# Filter-informed Spectral Graph Wavelet Networks for Multiscale Feature Extraction and Intelligent Fault Diagnosis

Tianfu Li, *Graduate Student Member*, *IEEE*, Chuang Sun, *Member*, *IEEE*, Olga Fink, *Member*, *IEEE*, Yuangui Yang, Xuefeng Chen, *Senior Member*, *IEEE*, Ruqiang Yan, *Fellow*, *IEEE*

*Abstract*—Intelligent fault diagnosis has been increasingly improved with the evolution of deep learning (DL) approaches. Recently, the emerging graph neural networks (GNNs) have also been introduced in the field of fault diagnosis with the goal to make better use of the inductive bias of the interdependencies between the different sensor measurements. However, there are some limitations with these GNN-based fault diagnosis methods. First, they lack the ability to realize multiscale feature extraction due to the fixed receptive field of GNNs. Secondly, they eventually encounter the over-smoothing problem with increase of model depth. Lastly, the extracted features of these GNNs are hard to understand owing to the black-box nature of GNNs. To address these issues, a filter-informed spectral graph wavelet network (SGWN) is proposed in this paper. In SGWN, the spectral graph wavelet convolutional (SGWConv) layer is established upon the spectral graph wavelet transform, which can decompose a graph signal into scaling function coefficients and spectral graph wavelet coefficients. With the help of SGWConv, SGWN is able to prevent the over-smoothing problem caused by long-range low-pass filtering, by simultaneously extracting low-pass and band-pass features. Furthermore, to speed up the computation of SGWN, the scaling kernel function and graph wavelet kernel function in SGWConv are approximated by the Chebyshev polynomials. The effectiveness of the proposed SGWN is evaluated on the collected solenoid valve dataset and aero-engine intershaft bearing dataset. The experimental results show that SGWN can outperform the comparative methods in both diagnostic accuracy and the ability to prevent over-smoothing. Moreover, its extracted features are also interpretable with domain knowledge.

## I. INTRODUCTION

AS a core part of Prognostics and Health Management (PHM), algorithms for intelligent fault diagnosis have been broadly used to enable an effective fault detection and isolation in rotating machinery such as aero-engines, helicopters, high-speed trains, and wind turbines [1]. By accurately diagnosing faults, equipment downtime and maintenance time can be significantly reduced, thereby improving production efficiency or providing useful guidance for decision-making of maintenance tasks [2, 3].

Both, machine learning (ML)- and deep learning (DL)- based methods have been successfully applied for fault diagnosis [4]. The traditional ML-based fault diagnosis methods use statistical analysis or signal processing methods to extract informative features. In the subsequent step fault classification is realized with machine learning approaches [5]. However, the feature extraction process of these traditional fault diagnosis methods is heavily dependent on the prior knowledge of domain experts, which is typically labor-intensive and empirical [6]. On the contrary, the DL-based fault diagnosis methods can effectively learn feature representations from the collected data and achieve accurate fault diagnosis in an end-to-end manner. Therefore, DL-based fault diagnosis methods can efficiently reduce the dependence on expert knowledge and improve the effectiveness of fault diagnosis [7-9]. For example, Li et al. [10] developed an end-to-end adversarial deep neural network to recognize unknown new faults. Wang et al. [11] proposed an improved convolutional neural network for multiscale feature fusion and fault diagnosis of rotating machinery. Huang et al. [12] proposed a wavelet packet decomposition-informed deep learning framework that obtained promising results for fault diagnosis of wind turbine gearboxes.

Although these DL-based methods have obtained impressive results for the fault diagnosis of rotating machinery, they may still leave the interplay between sensors measurements unexplored, because they can only achieve non-linear mapping and transformation in the Euclidean space [13]. Recently, graph neural networks (GNNs) have attracted a lot of attention due to their ability to learn the underlying dynamics of physical systems. They are able to model the interdependent correlations of data by aggregating the node features along the connecting edges between the vertices [14]. Early implementations of GNN architectures were built upon recurrent neural networks [15]. However, the most widely used GNNs today have been graph convolutional neural networks (GCNs). At present, there are two main streams for the current GCN models, i.e., spatial GCNs and spectral GCNs. The spatial GNNs directly define the graph convolution in the spatial domain by calculating the

This work was supported by Natural Science Foundation of China (No. 52175116), Major Research Program of Natural Science Foundation of China (No. 92060302) and the Fundamental Research Funds for the Central Universities. The work of Tianfu Li is also supported by China Scholarship Council (CSC) for one year's study at the École Polytechnique Fédérale de Lausanne (EPFL). (*Corresponding author: Chuang Sun*)

Tianfu Li, Chuang Sun, Yuangui Yang, Xuefeng Chen and Ruqiang Yan are with School of Mechanical Engineering, Xi'an Jiaotong University, Xi'an, Shaanxi 710049, China. Besides, Tianfu Li is also currently a visiting Ph.D. student of Laboratory of Intelligent Maintenance and Operations Systems, EPFL. (e-mail: litianfu@stu.xjtu.edu.cn\tianfu.li@epfl.ch, ch.sun@xjtu.edu.cn, yangyuangui@stu.xjtu.edu.cn, chenxf@xjtu.edu.cn, yanruqiang@xjtu.edu.cn)

Olga Fink is the Chair of Intelligent Maintenance and Operation Systems, EPFL, Lausanne 1015, Switzerland. (e-mail: olga.fink@epfl.ch)

weighted average of all vertices in its neighborhood [16-18]. While the theory of spectral GCNs is more solid, they define the graph convolution by utilizing the graph Fourier transform (GFT) [19] in graph signal processing. In practice, they first project the graph signals from the spatial domain to the spectral domain, and then amplify or attenuate the frequency components of interest with a spectral filter [20, 21], thereby, the signal frequencies of graph data are explored.

As the effectiveness of GCNs has been demonstrated in other application fields, GNNs have also been introduced to the field of fault diagnosis [22, 23]. For instance, Zhao et al. [24] developed a deep belief network-based graph convolution for the fault diagnosis of electromechanical systems. Zhou et al. [25] proposed a feature learning framework based on the dynamic graph for the fault diagnosis of rotating machinery under noisy working conditions. Zhao et al. [26] enhanced GCN with multiscale feature extraction and node attention and realized the fault diagnosis of a planar parallel manipulator. Although GCN-based fault diagnosis methods have recently made a considerable progress, they will still suffer from the following limitations due to the inheritance mechanism of GCNs:
1) Lack of the ability to realize multiscale feature extraction, due to the fixed receptive field of GCNs.
2) Lack of the ability to alleviate the over-smoothing problem as the depth of the network increases.
3) Lack of physical interpretability of the extracted features, due to the black box nature of GCNs.

Models with multiscale feature extraction ability can learn features in different resolutions and by that obtain more informative feature representations [27]. However, most GCNs only have a fixed receptive field, which can only aggregate information from one-hop neighbors by default [28]. Although high-order neighbor information can be explored by stacking multiple layers, the resulting deep GCN will have an over-smoothing problem due to the long-term low-pass filtering of GCN [29]. This makes the learned node representation of connected nodes indistinguishable. To increase the receptive field of GCNs, some works stacked multiple graph neurons with different scales to widen the network instead of increasing the depth of the network [30]. Nevertheless, the computational complexity of such GCNs is proportional to the number of scales stacked. Moreover, due to the requirements of trustworthy artificial intelligence, an important task in PHM is also to improve the physical interpretability of GCNs and their extracted features.

To enhance the physical interpretability of ML models, the concept of physics-informed ML [31] has recently been proposed, which refers to a class of methods that integrate domain constraints and underlying physics-based equations into ML models [32, 33]. Physics-informed ML has also been recently applied to fault diagnosis problems [34]. For example, Xin et al. [35] proposed a DL framework for the fault diagnosis of high-speed trains, in which the input signals were preprocessed with a logarithmic Short-time Fourier. To mimic wavelet transform analysis and combine the interpretability of signal processing with the learning capability of deep learning, Michau et al. [36] developed an unsupervised DL framework by integrating the process of wavelet transform into the model design. Gaetan et al [37] proposed a learnable wavelet packet transforms enabled deep learning framework, which can learn features automatically from data and extract the important time-frequency information of the dataset. However, the research studies on interpretable GCN-based intelligent fault diagnosis methods are still on-going and not fully explored.

To overcome the limitations of GCNs and achieve reliable fault diagnosis, we propose a filter-informed spectral graph wavelet network (SGWN) in this work. In SGWN, the proposed spectral graph wavelet convolution (SGWConv) layer is established upon the spectral graph wavelet transform, which can decompose the graph signal into a group of wavelet coefficients with one low-pass filter and multiscale band-pass filters. With help of SGWConv, the multiscale features of high-order neighbors can be captured. Thereby, the learned feature representations become more discriminative. Since the proposed SGWN incorporates the domain knowledge of graph signal processing in the model construction, the transparency of the model is improved, while the squared envelope spectrum is used to help understand its extracted features. To accelerate the computation of SGWN, the scaling and graph wavelet kernel functions of SGWConv are approximated by the $K$-order Chebyshev polynomials. The main contributions of this work are summarized as follows:
1) The filter-informed SGWConv layer is established based on the spectral graph wavelet transform, which can realize multiscale feature extraction by decomposing the graph signal into wavelet coefficients with one low-pass filter and multiscale band-pass filters.
2) An interpretable SGWN is built up by stacking multiple SGWConv layers to learn discriminative fault-related features, and its computation speed is accelerated by using Chebyshev polynomials to approximate the scaling and graph wavelet kernel functions in the SGWConv layer.
3) The over-smoothing problem is alleviated due to the ability of SGWN to extract multi-resolution features through the low-pass filter and multiscale band-pass filters rather than only perform long-range low-pass filtering.

The remainder of this paper is structured as follows. In Section II, the basic theory of graph Fourier transform and spectral graph wavelet transform are reviewed. The proposed SGWN is detailed in Section III. In Section IV, the performance of SGWN is evaluated on the collected solenoid valve dataset and aero-engine intershaft bearing dataset. In Section V, the ability of SGWN to overcome over-smoothing and the influence of hyperparameters and wavelet kernel functions are discussed. Finally, conclusions are made in Section VI.

## II. PRELIMINARIES

### A. Graph Fourier Transform based Graph Convolution

Given a graph G=(V, **A**, **X**), where V denotes the vertex set with |V|=$N$ and $N$ is the number of nodes, **A** is the adjacency matrix, in which $A_{i,j}$=1 only if the $i$-th and $j$-th node are connected, otherwise $A_{i,j}$=0. **X**=[**x**$_1$,**x**$_2$,…,**x**$_N$] with $\mathbf{X} \in \mathrm{R}^{N \times d}$ represents the node feature matrix and $d$ denotes the feature

dimension. The graph Laplacian matrix **L** of the graph is defined as **L**=**D**-**A**, where **D** is the degree matrix with $\mathbf{D}_{j,j}=\sum_i \mathbf{A}_{i,j}$. Since the graph Laplacian matrix **L** is symmetric, it has a set of orthonormal eigenvectors **U** which span $R^N$. Therefore, the matrix **L** can also be factored as $\mathbf{L}=\mathbf{U}\mathbf{\Lambda}\mathbf{U}^T$, where $\mathbf{\Lambda}=\text{dig}(\lambda_1,\lambda_1,\ldots,\lambda_N)$ is a diagonal matrix of eigenvalues $\lambda_N$, and $\mathbf{U}=[\mathbf{u}_1,\mathbf{u}_1,\ldots,\mathbf{u}_N]$ is the corresponding eigenvector matrix.

For a graph signal **x**, i.e., a node feature, the graph Fourier transform is defined as $\hat{\mathbf{x}}=\mathbf{U}^T\mathbf{x}$, in which its inverse form is denoted as $\mathbf{x}=\mathbf{U}\hat{\mathbf{x}}$. With graph Fourier transform (GFT) and based on the theory of convolution, the graph convolution operator $*_G$ can be mathematically denoted as [19]:

$$\mathbf{f} *_G \mathbf{x} = \mathbf{U}(\mathbf{U}^T\mathbf{f} \odot \mathbf{U}^T\mathbf{x}) \quad (1)$$

where $\odot$ is the Hadamard product and **f** denotes the convolution filter. The $\mathbf{U}^T\mathbf{f}$ can also be replaced by a diagonal filter matrix $\hat{\mathbf{f}}(\Theta)$. Then, the equation above becomes

$$\mathbf{f} *_G \mathbf{x} = \mathbf{U}\hat{\mathbf{f}}(\Theta)\mathbf{U}^T\mathbf{x} \quad (2)$$

where $\Theta$ is the weight parameter that should to be learned from the data and $\hat{\mathbf{f}}$ is the filter in the frequency domain.

### B. Spectral Graph Wavelet Transform

The spectral graph wavelet transform (SGWT) [38] can be deduced from the GFT and the continuous wavelet transform (CWT). The classical CWT can be generated by translating and dilating the mother wavelet $\psi(t)$ to build a continuous family of wavelets. The mother wavelet of CWT is defined as $\psi_{a,b}(t)=\frac{1}{a}\psi(\frac{t-b}{a})$, where $a>0$ and $b \in R$ are the scale parameter and translation parameter, respectively. In Fourier domain, the mother wavelet can also be reformulated as

$$\psi_{a,b}(t)=\frac{1}{2\pi}\int_{-\infty}^{\infty}\hat{\psi}(aw)e^{-iwb}e^{iwt}dw \quad (3)$$

where $\hat{\psi}$ is the wavelet function in frequency domain. The spectral graph wavelet can also be obtained in a similar way, where the spectral graph wavelet $\psi_{a,n}$ at scale $a$ and node $n$ is denoted as

$$\psi_{a,n} = \sum_{l=1}^{N} g(a\lambda_l)\mathbf{u}_l^T(n)\mathbf{u}_l = \mathbf{u}_l^T(n)g(a\mathbf{\Lambda})\mathbf{u}_l \quad (4)$$

where $g(\cdot)$ is the kernel function of the spectral graph wavelet that corresponds to the discrete scale $a=(a_1,a_2,\ldots,a_J)$, and $J$ denotes the decomposition scale. With the spectral graph wavelets, the graph wavelet coefficients $\mathbf{WT_x}$ can be defined as the inner product between the graph signal **x** and graph wavelets, which is expressed as

$$\mathbf{WT_x}(a,n) = \langle \psi_{a,n},\mathbf{x} \rangle = \sum_{l=1}^{N} g(a\lambda_l)\hat{\mathbf{x}}(l)\mathbf{u}_l(n) \quad (5)$$

where $\hat{\mathbf{x}}(l)=\mathbf{u}_l^T\mathbf{x}$ is the $l$-th component of GFT $\hat{\mathbf{x}}$.

The graph wavelet operator $g(a\lambda_l)$ works as a group of band-pass filters, while the scaling function in SGWT performs as a low-pass filter for the residual low-frequency component of the graph signal. Given a scaling kernel function $h$, the spectral scaling function $\phi_n$ and the corresponding scaling function coefficients at node $n$ can be analogously defined as

$$\phi_n = \sum_{l=1}^{N} h(\lambda_l)\mathbf{u}_l^T(n)\mathbf{u}_l = \mathbf{u}_l^T(n)h(\mathbf{\Lambda})\mathbf{u}_l \quad (6)$$

$$\mathbf{S_x}(n) = \langle \phi_n,\mathbf{x} \rangle = \sum_{l=1}^{N} h(\lambda_l)\hat{\mathbf{x}}(l)\mathbf{u}_l(n) \quad (7)$$

where $\mathbf{S_x}$ are the scaling function coefficients

## III. SPECTRAL GRAPH WAVELET NETWORKS

The proposed spectral graph wavelet convolution layer is described in detail in this section, including its polynomial approximation and kernel function design. In addition, a spectral graph wavelet network is also built up for intelligent fault diagnosis.

### A. The Filter-informed Spectral Graph Wavelet Convolution

As elaborated before, the core of SGWT is the design of a low-pass filter and multiscale band-pass filters in discretized scale. This definition shows that SGWT can achieve multiscale analysis by calculating the inner product between the graph signal and the band-pass filters. Therefore, by denoting the SGWT operator as **W**, the SGWT of a graph signal **x** can be expressed as

$$\mathbf{Wx} = [\phi, \psi_{a_1},\ldots,\psi_{a_J}]\mathbf{x}$$
$$= [\mathbf{U}h(\mathbf{\Lambda})\mathbf{U}^T\mathbf{x}, \mathbf{U}g(a_1\mathbf{\Lambda})\mathbf{U}^T\mathbf{x},\ldots, \mathbf{U}g(a_J\mathbf{\Lambda})\mathbf{U}^T\mathbf{x}]^T \quad (8)$$

While the Laplacian matrix **L** admits the eigenvalue decomposition $\mathbf{L}=\mathbf{U}\mathbf{\Lambda}\mathbf{U}^T$. Consequently, the equation above can be rewritten as

$$\mathbf{Wx} = [h(\mathbf{L})\mathbf{x}, g(a_1\mathbf{L})\mathbf{x},\ldots, g(a_J\mathbf{L})\mathbf{x}]^T \quad (9)$$

where the SGWT operator $\mathbf{W} = [h(\mathbf{L}), g(a_1\mathbf{L}),\ldots,g(a_J\mathbf{L})]^T \in R^{N\times(J+1)}$ consists of one scaling kernel function and $J$ wavelet kernel function at different scales.

By utilizing the SGWT operator **W** to replace the Fourier basis $\mathbf{U}^T$ in Eq. (2), we can obtain the expression for the filter-informed spectral graph wavelet convolution (SGWConv) as

$$\mathbf{f} *_G \mathbf{x} = \mathbf{W}^T\hat{\mathbf{f}}(\Theta)\mathbf{Wx} \quad (10)$$

where $\hat{\mathbf{f}}(\Theta)$ is a learnable diagonal filter matrix in wavelet domain, and $\mathbf{W}^T$ presents the inverse SGWT operator.

An illustration of the proposed SGWConv layer is depicted in Fig. 1. As can be seen, the SGWConv layer will decompose the input graph signal into scaling function coefficients and several scale-dependent wavelet function coefficients by utilizing the designed filters, thereby allowing for the multiscale feature extraction. After that, by passing through a filter matrix and calculating the inverse SGWT transform, the final feature representation can be obtained.

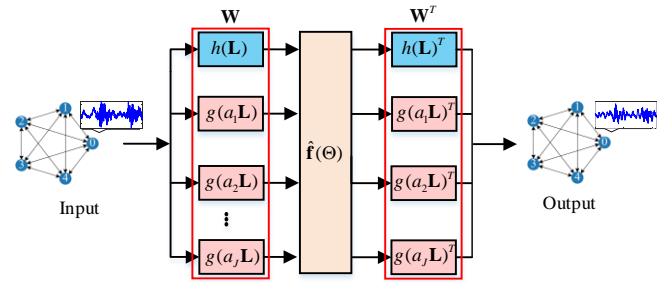

Fig. 1. The detailed process of SGWConv layer.

## B. Polynomial approximation of SGWConv Layer

In the SGWConv layer, to obtain the SGWT operator **W**, we need to calculate the scaling kernel function $h(\mathbf{L})$ and the graph wavelet kernel function $g(a_J\mathbf{L})$. Calculating $g(a_J\mathbf{L})$ corresponds to calculating $g(a_J\lambda_l)$ according to the properties of matrix functions. This would result in a high computational complexity of $O(N^3(J+1))$. Therefore, to achieve a better computational efficiency, the $h(\mathbf{L})$ and $g(a_J\mathbf{L})$ operators are approximated by the Chebyshev polynomials.

First, for a function $z(t)$ in the functional space, according to the approximation theory, we can find a convergent Chebyshev series of it, which is denoted as

$$z(t) = \frac{1}{2}c_0 + \sum_{k=1}^{\infty} c_k T_k(t) \quad (11)$$

where $c_k = \frac{2}{\pi}\int_0^{\pi}\cos(k\theta)z(\cos(\theta))d\theta$, and $k \le K$ is the order of the Chebyshev polynomials. Besides, the Chebyshev polynomials on interval $[-1, 1]$ can be iteratively defined by $T_k(t)=2tT_{k-1}(t)-T_{k-2}(t)$ with $T_0(t)=1$ and $T_1(t)=t$.

Therefore, for the scaling kernel function $h(\lambda_l)$ and the wavelet kernel function $g(a_j\lambda_l)$ with $\lambda_l \in [0, \lambda_{max}]$ and $\lambda_{max}$ is the maximum eigenvalue. To obtain the corresponding Chebyshev polynomial approximation, we can shift the domain of $\lambda_l$ to the interval $[-1, 1]$. Now, the approximations of scaling and wavelet kernel functions can be expressed as (See in Appendix for the detailed derivation process.)

$$h(\mathbf{L}) = \frac{1}{2}\bar{c}_{0,0} + \sum_{k=1}^{K} \bar{c}_{0,k}\bar{T}_k(\mathbf{L}) \quad (12)$$

$$g(a_j\mathbf{L}) = \frac{1}{2}\bar{c}_{j,0} + \sum_{k=1}^{K} \bar{c}_{j,k}\bar{T}_k(\mathbf{L}) \quad (13)$$

where $\bar{T}_k(L) = 2(\frac{2}{\lambda_N}\mathbf{L}-\mathbf{I})T_{k-1}(\mathbf{L})-T_{k-2}(\mathbf{L})$ represents the shifted Chebyshev polynomials, $\bar{c}_{0,k} = \frac{2}{\pi}\int_0^{\pi}\cos(k\theta)h(\frac{\lambda_{max}(\cos(\theta)+1)}{2})d\theta$ and $\bar{c}_{j,k} = \frac{2}{\pi}\int_0^{\pi}\cos(k\theta)g(\frac{a_j\lambda_{max}(\cos(\theta)+1)}{2})d\theta$ with $1 \le j \le J$. $\mathbf{I} \in \mathbf{R}^{N\times N}$ is the identity matrix.

By approximating **W** with the truncated shifted Chebyshev polynomial, the total computational complexity becomes $O(KN^2(J+1))$. In general, the order of Chebyshev polynomials $K$ is much smaller than $N$, i.e., $K \ll N$. Therefore, the computational complexity of SGWConv is significantly reduced through this polynomial approximation.

## C. Kernel Function Design for SGWConv Layer

As mentioned above, the appropriate wavelet kernel function and scale kernel function need to be defined in the SGWT operator **W**. It is then followed by the polynomial approximation. The kernel function for the SGWConv layer should satisfy the following constraints [39]

$$\begin{cases} g(0) = 0, \lim_{\lambda\to\infty} g(\lambda) = 0 \\ h(0) > 0, \lim_{\lambda\to\infty} h(\lambda) = 0 \end{cases} \quad (14)$$

In this paper, the most widely used Mexican hat wavelet is selected as the kernel function of the SGWConv layer. The Mexican hat wavelet kernel function and its corresponding scaling kernel functions are defined as proposed in [40]:

$$\begin{cases} g(t_j\lambda) = t_j\lambda\exp(-t_j\lambda) \\ h(\lambda) = \gamma\exp(-(\frac{Q\lambda}{0.6\lambda_{max}})) \end{cases} \quad (15)$$

where the scale value $t \in \{a|a_1, a_2,\ldots,a_J\}$ is determined by the maximum Laplacian eigenvalue $\lambda_{max}$ and the designed hyperparameter $Q$, in which the lower bound of the Laplacian eigenvalue is set to $\lambda_{min} = \lambda_{max}/Q$. Therefore, the minimum and maximum scales are defined as $a_1 = 1/\lambda_{max}$ and $a_J = 2/\lambda_{min}$, respectively. Then, we can set the intermediate scale to be decreasing logarithmically, so that $a_j = a_1(\frac{a_J}{a_1})^{\frac{j-1}{J-1}}$ with $1 \le j \le J$. Besides, we take $Q$ equals 2 here and the parameter $\gamma$ of the scaling kernel function is determined as $\gamma = g_{max}$, which equals to $e^{-1}$.

In order to give a basic understanding of the Mexican hat wavelet kernel function and its corresponding scaling kernel functions, we plot the frequency responses of these filters in Fig. 2. It can be observed that these curves clearly display the characteristics of low-pass and band-pass filters of the scaling kernel and wavelet kernel functions, i.e., the scaling kernel approaches 0 as $\lambda\to\infty$, while each of the wavelet kernel approaches 0 as $\lambda\to 0$.

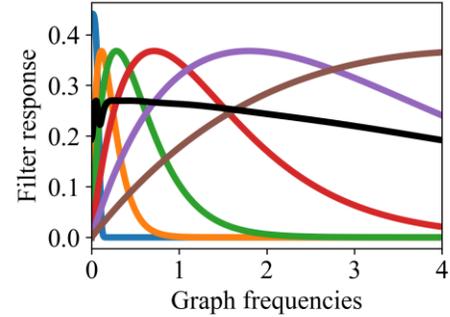

Fig. 2. Illustration of Mexican hat wavelet kernel function $\{g(t_j\lambda)\}_{j=1}^{J}$ with $J=5$ (the orange, green, red, pink, and brown curve, respectively) and its scaling kernel function (the blue curve), while the black curve is the sum of squares of these filters.

## D. Spectral Graph Wavelet Network based Intelligent Fault Diagnosis

The spectral graph wavelet network (SGWN) is obtained by stacking two SGWConv layers for multiscale feature extraction and two fully connected layers for fault classification. The architecture of the designed SGWN is shown in Fig. 3, and the formulation of each layer is listed below.

Firstly, the two SGWConv layers of SGWN are defined as follows:

First SGWConv layer: $\mathbf{H}_1 = \text{ReLu}(\mathbf{W}^T\hat{\mathbf{f}}(\mathbf{\Theta}_1)\mathbf{W}\mathbf{X})$ (16)

Second SGWConv layer: $\mathbf{H}_2 = \text{ReLu}(\mathbf{W}^T\hat{\mathbf{f}}(\mathbf{\Theta}_2)\mathbf{W}\mathbf{H}_1)$ (17)

where **X** and **H** denote the input and learned node representations, respectively. $\hat{\mathbf{f}}(\mathbf{\Theta}_1)$ and $\hat{\mathbf{f}}(\mathbf{\Theta}_2)$ are the learnable weight matrices of the two SGWConv layers, and ReLu(·) is the activation function.

Subsequently, to obtain the entire representation of the graph, the output from the second SGWConv layer will be fed into a readout layer, in which the graph representation is defined as

$$\overline{\mathbf{H}} = \frac{1}{N}\sum_{i=1}^{N}\mathbf{H}_2 \qquad (18)$$

where $\overline{\mathbf{H}}$ is the final graph representation.

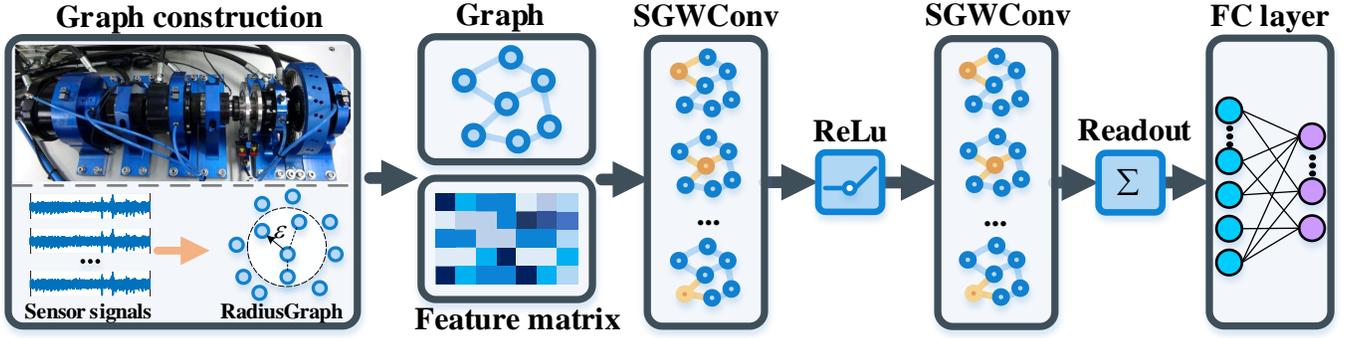

Fig. 3. The framework of SGWN for intelligent fault diagnosis.

Finally, the obtained graph representations are inputted into two fully connected (FC) layers for fault classification, and the predicted probability $\overline{p}$ of a fault mode is expressed as

$$\overline{p} = \text{softmax}(\text{FC}(\overline{\mathbf{H}})) \qquad (19)$$

where softmax(·) is used to normalize the output of the FC layers to a probability distribution over predicted output classes.

For model training, the cross-entropy (CE) loss is adopted as the loss function, which is expressed as

$$\text{CE} = -\frac{1}{M}\sum_{i}\sum_{c=1}^{C} y_{ic}\log(\overline{p}_{ic}) \qquad (20)$$

where $M$ and $C$ represent the number of graphs and fault classes, respectively; $p_{ic}$=0 if the label of graph $i$ is not equal to $c$, otherwise $p_{ic}$=1. We summarize the process described above in **Algorithm I**.

| **Algorithm I** SGWN based Intelligent fault diagnosis |
|---|
| INPUTS: |
| • $M$: Number of graphs; |
| • $G_m=(V_m, \mathbf{A}_m, \mathbf{X}_m)$: Graphs of a fault mode with label $y_m$, $1 \leq m \leq M$. |
| MODEL HYPERPARAMETERS: |
| • $J$: Number of decomposition scales for $\{a\|a_1,a_2,\ldots,a_J\}$; |
| • $K$: Order of Chebyshev polynomials. |
| SETPS: |
| 1) Select proper $h(\mathbf{L})$ and $g(a_J\mathbf{L})$ that satisfy Eq. (14); |
| 2) Approximate SGWT operator $\mathbf{W}$ for each input graph $G_m$ with scale $a$ using Eq. (9), Eq. (12), and Eq. (13); |
| 3) **For** model training **do** |
|    a) $\overline{p} \leftarrow$ SGWN($G_m$, $J$, $K$); |
|    b) CE ← $-\frac{1}{M}\sum_i \sum_{c=1}^C y_{ic}\log(p_{ic})$; |
|    c) update the weight matrices $\hat{\mathbf{f}}(\mathbf{\Theta})$ and parameters of FC layers with error backpropagation algorithm; |
| **End for** |
| OUTPUTS: |
| • Network parameters for SGWN; |
| • The predicted fault types of the test set. |

## IV. CASE STUDY

In this section, the effectiveness of the proposed SGWN is evaluated on two case studies: 1) the fault diagnosis of a solenoid valve and 2) fault diagnosis of an aero-engine intershaft bearing. Moreover, the de-noising ability and interpretability of SGWN are discussed.

### A. Fault Diagnosis of Solenoid Valve of Fuel Control System

#### 1) Data Description

As the solenoid valve is a key component of the automatic control of the fuel control system, its failure can cause the system function either incorrectly or inefficiently. To collect failure data of solenoid valves, an experiment was conducted on the fuel control system, the test bench of which is shown in Fig. 4(a), and its detailed schematic diagram is illustrated in Fig. 4(b). As can be seen, this fuel control system mainly consists of a plunger pump, a motor, a solenoid reversing valve, a proportional relief valve, and a solenoid valve.

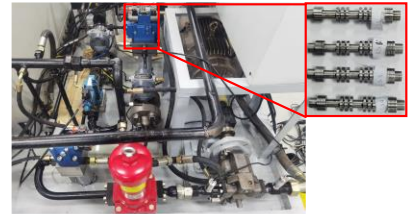

(a) Fuel Control System

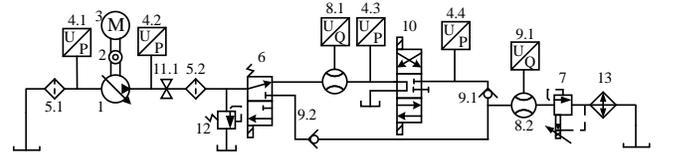

1. Plunger pump; 2. Torque Tachometer; 3. Motor; 4. Pressure Sensor; 5. Oil filter; 6. Solenoid reversing valve; 7. Proportional relief valve; 8. Flowmeter; 9. Check valve; 10. Solenoid valve; 11. Ball valve; 12. Safety valve; 13. Cooler.

(b) Schematic diagram

Fig. 4. Fuel control system test bench.

TABLE I
DESCRIPTION OF THE FAULT SATE OF SOLENOID VALVE

| System health state | Description |
|---|---|
| I | Solenoid valve spool is pulled by 2mm |
| II | Solenoid valve spool is pulled by 3mm |
| III | Solenoid valve spool is pulled by 4mm |
| IV | Solenoid valve spool is pulled by 7mm |
| V | Solenoid valve spool wear 1mm |
| VI | Solenoid valve spool wear 2mm |
| VII | Solenoid valve spool wear 3mm |
| VIII | Solenoid valve spool wear 4mm |
| IX | Normal state |

In this experiment, we conduct a fault simulation experiment on the solenoid valve. We prefabricate four different degrees of the strain and wear failure of the solenoid valve spool, respectively, in which the spool strain failure is shown in the red box in Fig. 4(a). In addition to the normal state, a total of nine sets of data are collected in the experiment, and the fault states are illustrated in Table I. During the experiment, 10 sensors are used to collect data, including the input and output pressure signals of the plunger pump, the speed and torque signals of the motor, the input and output pressure, and the input and output flow signals of the solenoid valve, and two acceleration sensors are used to collect the vibration signals of the solenoid valve. The sampling frequency is 20480 Hz, the rotating speed of the motor is 2000 rpm and the oil pressure is 8 MPa.

*2) Implementations Details*
*a) Data Preprocessing and Graph Data Construction*

In this experiment, the measurement values of each sensor signal are normalized with Max-min normalization. After the normalization step, we can build the sensor network, where each sensor represents a node, and generate spatial-temporal graphs using this sensor network. While for the sensor network construction, the KNNGraph and RadiusGraph are two commonly used methods [22]. In this paper, we adopt the RadiusGraph to construct the sensor network, which can be defined by

$$AN(\mathbf{x}_i) = \varepsilon\text{-radius}(\mathbf{x}_i, \mathbf{X}) > \varepsilon \quad (21)$$

where AN($\mathbf{x}_i$) denotes the adjacent nodes of the *i*-th sensor measurements $\mathbf{x}_i$, and $\varepsilon$-radius($\mathbf{x}_i$, $\mathbf{X}$) calculates the cosine similarity between each sensor measurements and returns the adjacent nodes only if the cosine similarity is greater than $\varepsilon$.

After obtaining the sensor network, a sidling window with length *d* (i.e., the feature dimension) is applied to truncate the collected signal, and the truncated signal measurements are considered as the node features of the graph. With the signal length *P*, $M=\lceil P/d \rceil$ graphs can be obtained. Moreover, we can obtain a group of spatial-temporal graphs with the same graph structure but different time-stamp node features.

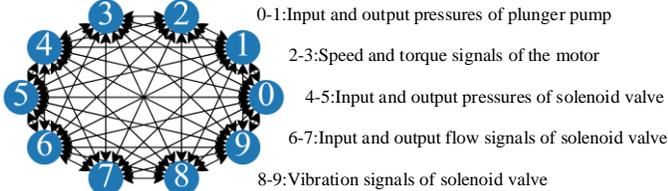

Fig. 5. The constructed sensor network.

In this experiment, the constructed sensor network of the normal state is shown in Fig. 5. In graph data construction, a length of the sliding window with length 1024 is applied to truncate the signals without overlap. For each failure mode, we obtain 1000 samples with each of them representing a graph, in which we randomly select 80% of samples for model training and the rest 20% for model testing. Therefore, a total of 72000 graphs are in the training set and 1800 graphs in the test set.

*b) Experimental Setups and Comparative Methods*

In order to evaluate the effectiveness of the proposed SGWN, 10 state-of-the-art methods are implemented as the comparative methods, which are detailed in the Table II. In order to ensure a fair comparison, the number of layers of all the models is kept the same as that of SGWN as in Table III.

In the experiment, the batch size is set to 100, the learning rate is initialized as 0.01, and a learning rate decay strategy is adopted. All model is implemented on the Pytorch framework with the Windows 10 operating system, and hardware with Intel i7-9700K CPU and a GTX2080Ti GPU. For model training, each model is trained for 100 epochs with the Stochastic Gradient Descent (SGD) optimizer. Besides, for SGWN, the scale parameter and the order of Chebyshev polynomials are both set to 2. The overall classification accuracy is leveraged as the evaluation metric, and to reduce the randomness of the results, each experiment is repeated for 5 times and the average value of these results is used as the final result.

TABLE II
THE DESCRIPTION OF THE COMPARATIVE METHODS

| Model type | Description |
|---|---|
| Spectral GNN | **GCN** [20] is the most famous spectral-GNN, which is defined on the graph Fourier transform. **SGCN** [21] simplifies GCN and has less learnable parameters. Here, it has two graph convolutional (GConv) layers and one readout layer. **MRFGCN** [30] is a multiscale GNN, which can realize multiscale feature extraction by stacking GConv layers with different receptive fields. Here, the stacked two GConv layers both contain three different receptive fields. |
| Spatial GNN | **GraphSage** [16] is the earliest spatial GNN, which uses the AGGREGATE(·) and CONCAT(·) functions for node feature updating. **GIN** [18] is a spatial-GNN, which uses sum aggregator for feature aggregation and an MLP for feature updating. **GAT** [17] achieves feature updating with attention mechanism that calculates the attention coefficient of each node. Single head attention is adopted here. **GateGCN** [15] is an RNN like model, which can model the sequences of graph-structured data. |
| Other DL models | **CNN** and **BiLSTM.** The implemented CNN consists of two convolutional layers with kernel size is 3 and two pooling layers [41]. The BiLSTM contains two LSTM cells and the length of the output is 1024. **Transformer** is a new elegant framework, which builds upon the multi-head self-attention mechanism and completely abandons recurrence and convolutions [42]. |

TABLE III
THE DETAILED MODEL STRUCTURE

| Layer | Input size | Output size |
|---|---|---|
| SGWConv1 | $B \times N \times 1024$ | $B \times N \times 1024$ |
| BatchNorm | $B \times N \times 1024$ | $B \times N \times 1024$ |
| ReLU | $B \times N \times 1024$ | $B \times N \times 1024$ |
| SGWConv2 | $B \times N \times 1024$ | $B \times N \times 1024$ |
| BatchNorm | $B \times N \times 1024$ | $B \times N \times 1024$ |
| ReLU | $B \times N \times 1024$ | $B \times N \times 1024$ |
| Readout | $B \times N \times 1024$ | $B \times 1024$ |
| FC1 | $B \times 1024$ | $B \times 512$ |
| FC2 | $B \times 512$ | $B \times C$ |

\**B* is the batchsize, *N* and *C* are the number of nodes and classes, respectively.

*3) Diagnostic Results*

The experimental results are listed in Table IV and the diagnostic accuracy of each trial is shown in Fig. 6. In Table IV, we also report the minimum accuracy (Min-acc), maximum accuracy (Max-acc), and the average training and testing times.

It can be observed from these results that the best diagnostic accuracy is achieved by the proposed SGWN with 97.78%. In terms of average accuracy, SGWN achieves an improvement between 1% for BiLSTM and 17% for the SGCN. In addition, it shows that the standard deviation (STD) of SGWN is much smaller compared to other methods, which illustrates the stability of SGWN. The training and testing time required for SGWN is greater than that of single-scale models such as GCN, but less than that of multi-scale models such as MRFGCN, in which the extra time it takes is mainly due to the calculation of the spectral wavelet coefficients.

TABLE IV
THE DIAGNOSTIC ACCURACY (%) OF SOLENOID VALVE DATASET

| Models | Min-acc | Max-acc | Avg-acc ±STD | Training time(s) | Testing time (s) |
|---|---|---|---|---|---|
| GCN | 83.61 | 88.67 | 87.17±2.02 | 137 | 34 |
| SGCN | 67.72 | 90.28 | 80.47±8.33 | 134 | 34 |
| MRFGCN | 93.5 | 95.83 | 94.68±1.02 | 238 | 60 |
| GraphSage | 81.78 | 96.39 | 92.87±6.24 | 127 | 32 |
| GIN | 61.67 | 82.83 | 72.43±8.94 | 128 | 32 |
| GAT | 83.83 | 91.72 | 88.77±2.99 | 165 | 41 |
| GateGCN | 86.22 | 92.06 | 89.72±2.41 | 167 | 42 |
| CNN | 88.61 | 95.17 | 92.07±2.46 | 127 | 32 |
| BiLSTM | 95.67 | 97.18 | 96.37±0.46 | 257 | 64 |
| Transformer | 57.33 | 61.17 | 59.45±1.37 | 204 | 51 |
| SGWN | **97.22** | **97.78** | **97.47±0.22** | 227 | 57 |

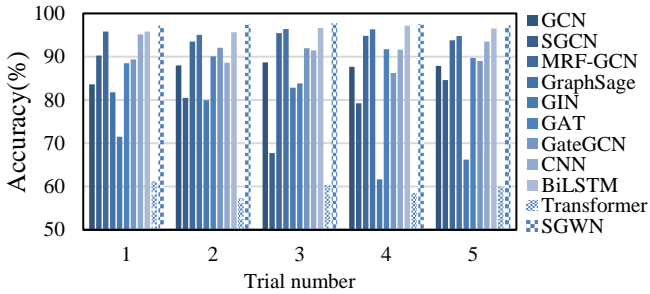

Fig. 6. The diagnostic accuracy of solenoid valve dataset under each trail.

*4) The De-noising Ability of SGWN*

In practice, the collected vibration signals of machinery often contain interference components such as environmental noise. Therefore, in order to evaluate the de-noising ability of SGWN, we add Gaussian white noise with different signal-to-noise ratios (SNR) to the raw signal, where the SNR is denoted as

$$\text{SNR}=10\log\left(\frac{P_s}{P_n}\right) \quad (22)$$

where $P_s$ and $P_n$ are the power of signal and noise, respectively. Due to the lower value of the SNR, more noise will be contained in the processed signal. We add three different intensities of noise to the raw signal with SNR of 0dB, -3dB, and -5dB, in which the SNR equal to -5dB means that the energy of the noise is larger than three times the energy of the raw signal. The diagnostic results are shown in Table IV and Fig. 7.

TABLE V
THE DIAGNOSTIC ACCURACY (%) OF SOLENOID VALVE DATASET UNDER DIFFERENT SNR

| Models | -5dB | -3dB | 0dB |
|---|---|---|---|
| GCN | 43.52±8.11 | 43.98±3.57 | 58.34±4.01 |
| SGCN | 42.60±8.06 | 44.96±2.80 | 60.53±2.83 |
| MRFGCN | 77.86±6.91 | 84.50±4.54 | 90.44±3.58 |
| GraphSage | 77.83±6.07 | 84.48±3.78 | 89.63±2.03 |
| GIN | 40.39±3.69 | 50.53±1.81 | 63.56±1.79 |
| GAT | 46.07±8.19 | 48.77±6.89 | 61.90±6.34 |
| GateGCN | 49.16±5.89 | 54.69±9.91 | 65.79±5.39 |
| CNN | 37.36±2.65 | 62.96±3.09 | 68.25±2.25 |
| BiLSTM | 74.25±3.19 | 79.11±2.94 | 84.43±3.84 |
| Transformer | 31.10±3.09 | 37.45±3.18 | 43.28±1.42 |
| SGWN | **82.07±2.89** | **85.54±2.10** | **91.81±1.22** |

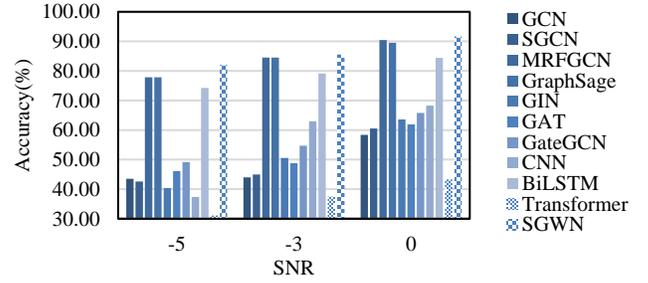

Fig. 7. The diagnostic accuracy of each model under different SNR.

It can be seen from these results that with the decrease of the SNR, the performance of each model decreases, while the standard deviation of each model increases. Compared to the other methods, even at -5dB SNR, the proposed SGWN can still obtain the best diagnostic results with the smallest standard deviation. Consequently, the results shown in Table V and Fig. 7 demonstrate that the de-noising ability of SGWN outperforms the comparative methods.

In order to show the impact of noise on model performance in a more intuitive way, we use *t*-SNE to visualize the prediction results of SGWN under various noise conditions, which is illustrated in Fig. 8. It can be found that as the noise increases the learned feature representations tend to increasingly become indistinguishable. Especially, when SNR=-5, the learned representations of some fault modes become very difficult to distinguish.

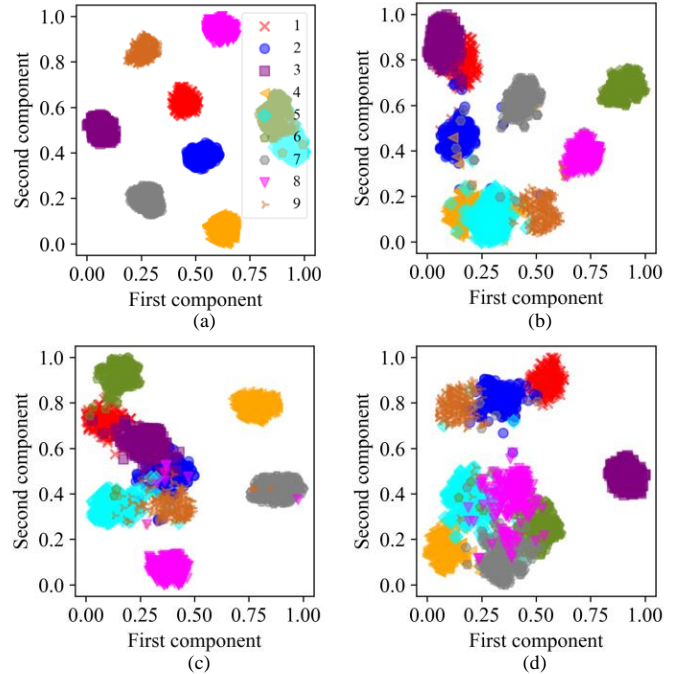

Fig. 8. The t-SNE visualization of the results at different SNR. (a) None; (b) SNR=0; (c) SNR=-3; (d) SNR=-5.

### B. Fault diagnosis of Aero-engine Intershaft Bearing

*1) Data Description*

Aero-engines usually work under harsh conditions. Due to the safety-critical operation, the failure of intershaft bearings can result in catastrophic accidents. To collect the failure data of the aero-engine intershaft bearing, the experiment is implemented on the dual rotor test bench, as shown in Fig. 9. As can be seen, the dual rotor test bench mainly consists of a low-pressure (LP) shaft and a high-pressure (HP) shaft, where the LP shaft has the same axis as the HP shaft and the intershaft bearing is used to connect the two shafts. Moreover, two motors are utilized to drive the low-pressure shaft and the high-pressure shaft to rotate at different speeds.

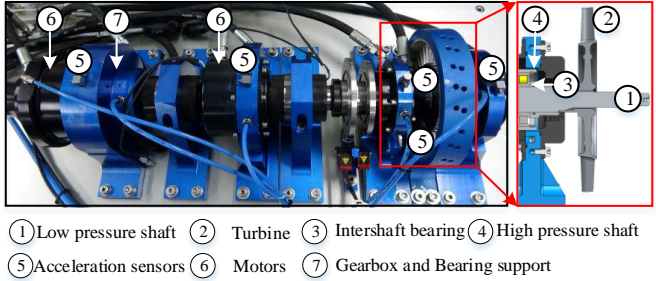

① Low pressure shaft ② Turbine ③ Intershaft bearing ④ High pressure shaft
⑤ Acceleration sensors ⑥ Motors ⑦ Gearbox and Bearing support

Fig. 9. Dual rotor test bench.

In this experiment, the vibration signals from eight N205 bearings with non-uniform health conditions are collected, including bearing inner ring (IR) and rolling element (RF) fault with three different fault severities, and bearing outer ring (OR) fault with two different fault severities. Together with the normal state, it can be considered as a nine-class classification task. The synthetically induced bearing faults are shown in Fig. 10. During the experiment, the rotating speeds of the LP shaft and HP shaft are set to 4000 rpm and 8000 rpm. Five acceleration sensors are used to collect the vibration signals with a sampling frequency is 20480 Hz.

The data normalization and graph generation strategies are consistent with the previous experiments. This means that the sensor network is constructed based on the five acceleration sensor measurements. With such a sensor network, we can also obtain 1000 graphs for each fault mode. Therefore, there will be 7200 graphs for model training and 1800 graphs for the model testing.

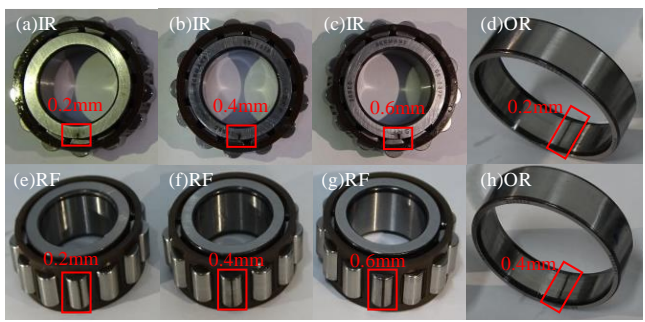

Fig. 10. Bearings with eight different fault modes.

*2) Diagnosis Results*

The experimental results are shown in Table VI and Fig. 11. Compared to the 10 comparison methods, SGWN obtains an overall gain of between 1.4% of BiLSTM and 43.3% of GCN in terms of the average accuracy. Moreover, SGWN can also achieve the smallest standard deviation among the ten comparative methods. The difference in training and testing time is consistent with the previous experiments. These experimental results further confirm the superiority of SGWN compared to other methods.

TABLE VI
THE DIAGNOSIS ACCURACY (%) OF INTERSHAFT BEARING DATASET

| Models | Min-acc | Max-acc | Avg-acc ±STD | Training time(s) | Testing time (s) |
|---|---|---|---|---|---|
| GCN | 55.5 | 57.33 | 56.68±0.77 | 137 | 34 |
| SGCN | 57.78 | 59.56 | 58.46±0.71 | 134 | 34 |
| MRFGCN | 98 | 99.89 | 98.70±0.40 | 238 | 60 |
| GraphSage | 96.67 | 96.78 | 96.72±0.06 | 127 | 32 |
| GIN | 65.44 | 68 | 66.62±1.03 | 128 | 32 |
| GAT | 53.78 | 66.22 | 57.43±4.99 | 165 | 41 |
| GateGCN | 97.89 | 98.89 | 98.41±0.31 | 167 | 42 |
| CNN | 97.78 | 98.28 | 98.06±0.18 | 110 | 32 |
| BiLSTM | 96.67 | 99.78 | 98.57±1.09 | 257 | 64 |
| Transformer | 54.78 | 62.11 | 58.04±2.94 | 204 | 51 |
| SGWN | **99.89** | **100** | **99.94±0.04** | 227 | 57 |

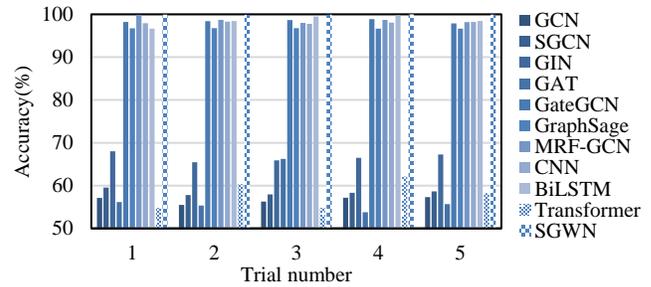

Fig. 11. The diagnosis accuracy of intershaft bearing dataset under each trail.

*3) Evaluate the Extracted Features of SGWN*

Although DL-based intelligent fault diagnosis models can achieve good diagnosis accuracy, their extracted features are typically difficult to be interpreted. To evaluate the content of the learned features, we process them with traditional signal processing approaches that the experts are familiar with in fault diagnostics. One of the indications that the extracted features contain the relevant information on the fault can be assessed by evaluating if the relevant fault characteristic frequencies ($f_r$) can be extracted from the features. It is typically used by domain experts to evaluate the signatures of the faults. If this information is missing in the extracted features, experts will not be able to interpret the obtained results [31].

With this basic concept, we perform squared envelope spectrum (SES) analysis [43] on the extracted features of SGWN at each scale and its output features to find fault-related frequencies to interpret the underlying fault. In this part, we take the bearing inner ring with a 0.4 mm fault severity as an example. The resulting fault characteristic frequency of this bearing can be calculated as $f_r$=518.03Hz. Since the SGWN contains two SGWConv layers, we visualize the obtained multiscale features of the first acceleration sensor that is extracted by the first SGWConv layer. The extracted scaling function (SF) coefficients and spectral graph wavelet (SGW) coefficients are shown in Fig. 12(Left). Their corresponding SESs are shown in Fig. 12(Right).

As can be seen from these results, the SGWConv layer can effectively decompose the input signal into SF coefficients and multiscale SGW coefficients, where the SF coefficient has the largest magnitude and contains most of the energy of the signal (Interestingly, we find the same conclusion as Ref.[40]). Moreover, in the calculated SESs, the fault characteristic frequency of the bearing can be precisely located. This indicates that the extracted multiscale features can efficiently preserve the fault-related components of the signal. This also illustrates why the corresponding fault characteristic frequencies can be found in the output features of the SGWConv layer.

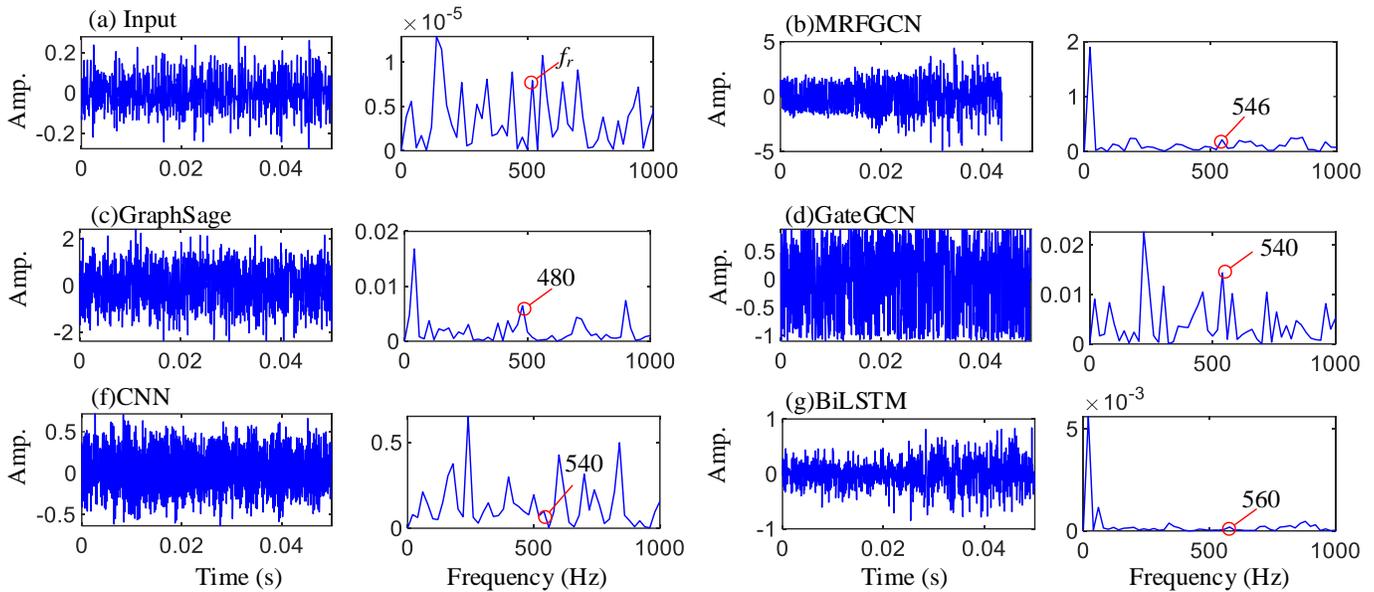

Fig. 13. The extracted features of the five best performing comparison methods and their corresponding spectra.

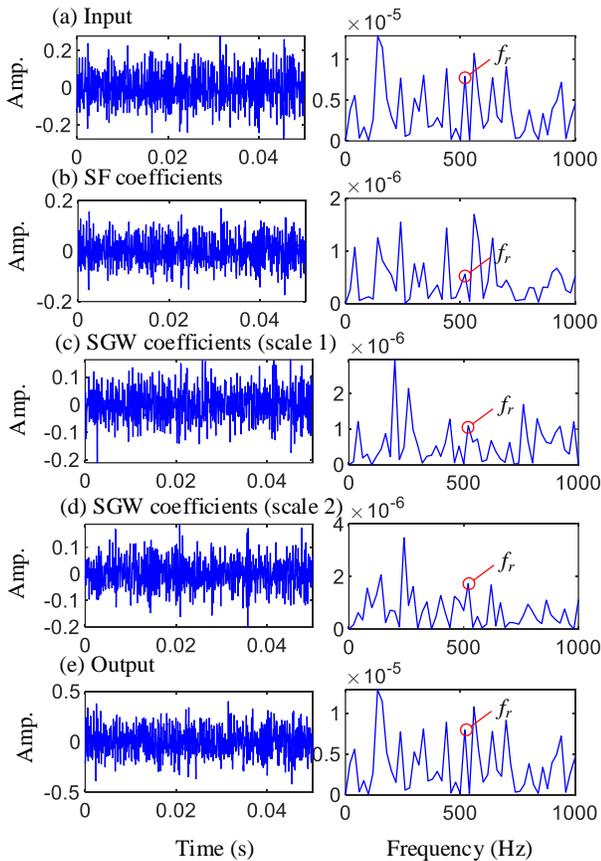

Fig. 12. Multi-scale features extracted from the input signal by SGWConv. **Left**: the extracted SF and SGW coefficients in the time domain; **Right**: the corresponding square envelope spectra.

Furthermore, to better understand which features are learned by the other comparison methods and to evaluate if they are similar to those obtained with the proposed methodology, we also perform envelope spectrum analysis on the output of the first layer of the five best performing comparison models in aero-engine intershaft bearing fault diagnosis. As can be seen from Fig. 13, although these models obtain very good diagnosis accuracy, their outputs not only deviate from the original input, but also lose the fault characteristic frequency. This is particularly true for especially the MRFGCN, GateGCN, and BiLSTM. This makes the extracted features more difficult to interpret by experts. Therefore, all these results indicate that the learned features of SWGN can be better understood by domain experts.

## V. DISCUSSION

In this section, the ability of SGWN to overcome the over-smoothing problem, and the influence of hyperparameters and wavelet kernel functions on model performance are discussed.

### A. The Ability of SGWN to Alleviate Over-smoothing Problem

The working principle of most GNNs can be summarized as a message-passing mechanism [44], which is equivalent to performing low-pass filtering on feature vectors. As has been investigated in Ref. [20, 29], with the increase of the model depth, the long-range low-pass filtering will lead to the GCNs becoming over-smoothed, thus making it difficult to distinguish between adjacent node representations. Therefore, to verify the ability of SGWN to avoid the over-smoothing problem, we increase the depth of the SGWN. To compare the performance of the SGWN, we select one of the most commonly used spectral GCN (i.e., GCN) and one spatial GCN (i.e., GAT) for comparison. The experimental results are shown in Fig. 14.

It can be found that with the increase of the model depths, the diagnosis results of each model gradually increased. When the model depth of GCN and GAT exceeds six layers, their performance begins to decrease, especially when the number of network layers reaches 20. They can only achieve a diagnosis accuracy of 67.83% and 30.39%, respectively. In contrast, the output of SGWN only results in slight fluctuations as the number of network layers increases, and can still achieve competitive diagnosis results when the model depth is 20 layers. These results show that the multiscale feature extraction achieved by the low-pass filter and band-pass filters can well preserve discriminative properties of features, making them distinguishable in deeper and more complex networks. In addition, in terms of the number of network layers, after referring to the suggestions given in Kipf et al. [20], we also recommend that the optimized layer number of SGWN should be no more than six to balance training time and model performance.

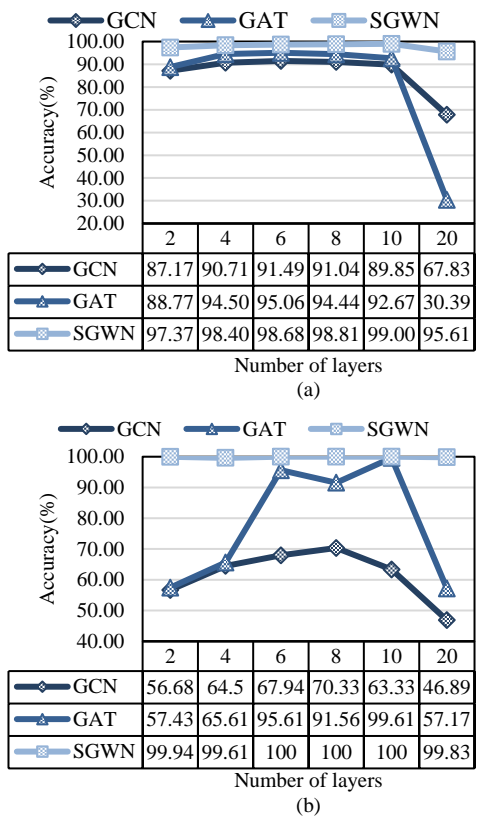

Fig. 14. The diagnosis accuracy at different model depths. (a) the solenoid valve dataset; (b): the intershaft bearing dataset.

### B. The Influence of the Model Hyperparameters

The decomposition scale and the order of the Chebyshev polynomial are two very important hyperparameters in the proposed SGWN. Moreover, the decomposition scale and order that are too large will increase the complexity of the model. Therefore, we set the initial value of the decomposition scale and order of the Chebyshev polynomial both to two. During the experiments, the decomposition scale is increased from 2 to 10 when the order of Chebyshev polynomials takes 2, and vice versa. The experimental results are shown in Fig. 15.

As can be seen from these results, as the number of scales increases, the diagnosis results of SGWN remain stable across all numbers of scales, while the training time increases linearly. These results indicate that SGWN is not sensitive to the number of decomposition scales, which is mainly attributed to the remarkable numerical stability of the Chebyshev polynomials. In addition, the model performance of SGWN is also relatively stable under different Chebyshev orders with a maximum deviation of 1%. Therefore, considering the training time, we recommend that the number of scales and the order of the Chebyshev polynomial both take two or three.

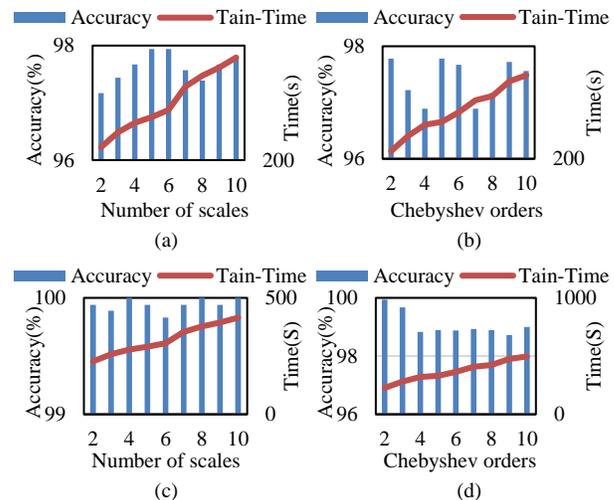

Fig. 15. The diagnostic results of SGWN with different hyperparameters. (a)~(b): the solenoid valve dataset; (c)~(d): the intershaft bearing dataset.

### C. The Influence of the Wavelet Kernel Function

As mentioned in Section III, many wavelet kernel functions have been developed in SGWT. In order to evaluate how the wavelet kernel functions influence the model performance, we construct two variants of SGWNs, namely SGWN-C and SGWN-H with cubic spline wavelet and heat kernels. The cubic spline wavelet has the same scaling function as the Mexican hat wavelet, while the heat kernel does not have the corresponding scaling kernel function. For convenience, we list the three wavelet kernel functions used here in Table VII, in which $\alpha$ and $\beta$ are both set to two, and the $\lambda_1$ and $\lambda_2$ of the cubic spline wavelet take one and two, respectively. To explore the reduced computational complexity of the Chebyshev polynomial approximation, we also add non-approximated versions of each SGWN. The experimental results are shown in Table VIII.

TABLE VII
THE FORMULA OF THE USED THREE WAVELET KERNELS

| Wavelet name | Wavelet kernel $g$ | Scaling kernel $h$ |
|---|---|---|
| Cubic spline [38] (SGWN-C) | $\begin{cases} \lambda_1^{-\alpha}\lambda^{\alpha}, \text{ for } \lambda < \lambda_1 \\ s(\lambda), \text{ for } \lambda_1 \leq \lambda \leq \lambda_2 \\ \lambda_2^{-\beta}\lambda^{\beta}, \text{ for } \lambda > \lambda_2 \end{cases}$ $s(\lambda)=-5+11\lambda-6\lambda^2+\lambda^3$ | $\gamma \exp(-(\frac{Q\lambda}{0.6\,\lambda_{\max}})^4)$ |
| Mexican hat [40] (SGWN) | $\lambda\exp(-\lambda)$ | $\gamma \exp(-(\frac{Q\lambda}{0.6\,\lambda_{\max}})^4)$ |
| Heat kernel [45] (SGWN-H) | $exp(-\lambda)$ | |

As can be seen from these results, the proposed SGWN-H and SGWN achieved competitive performances among the 10 comparative methods, while SGWN-C can only achieve a diagnosis accuracy of 53.32% and 94.61% on the solenoid valve and intershaft bearing dataset, respectively. The bad performance of SGWN-C on the solenoid valve dataset shows that the cubic spline wavelet with the piecewise kernel function is not suitable for processing the data consisting of low-frequency signals such as pressure and flow. Besides, compared to SGWN-H, the performance of SGWN is more stable and has a smaller standard deviation.

TABLE VIII
THE DIAGNOSIS ACCURACY OF THREE DIFFERENT SGWNs

| Type | Model | Solenoid valve dataset | Intershaft bearing dataset | Training time(s) |
|---|---|---|---|---|
| Approximate | SGWN-C | 53.32±2.3 | 94.61±1.2 | 271 |
|  | SGWN-H | 97.63±0.46 | 99.98±0.07 | 250 |
|  | SGWN | 97.37±0.27 | 99.94±0.04 | 227 |
| Non-approximated | SGWN-C | 61.67±1.33 | 95.32±0.72 | 307 |
|  | SGWN-H | 97.68±0.32 | 100±0.01 | 273 |
|  | SGWN | 97.51±0.14 | 100±0.01 | 255 |

In terms of training costs, it can be observed that the SGWN with Cubic spline wavelet took the longest training time, while the SGWN with Mexican hat wavelet took the least training time. Furthermore, we also find that the computational complexity of SGWN with polynomial approximation is greatly reduced. For SGWN-C, SGWN-H, and SGWN, the computational complexity after approximation is reduced by 28 to 36 seconds, respectively. Therefore, in terms of diagnosis results and computational complexity, it is suggested to choose the Mexican hat wavelet as the kernel function of SGWN.

## VI. CONCLUSION

To realize multiscale feature extraction, alleviate over-smoothing problems, and increase the physical interpretability of the current GCN-based fault diagnosis methods, this paper proposed a filter-informed SGWN. In SGWN, the SGWConv layer is established by incorporating the domain knowledge of SGWT, which can decompose the signal through one low-pass filter and the multiscale band-pass filters. Due to the SGWConv layer containing multiple band-pass filters, it can perform a multiscale feature extraction and alleviate the over-smoothing problem caused by long-range low-pass filtering. The effectiveness of the proposed SGWN has been verified on the collected solenoid valve dataset and aero-engine intershaft bearing dataset. The experimental results show that SGWN is physically interpretable and can also achieve the best results among the comparative methods. Besides, the ability to avoid over-smoothing problems is verified. Moreover, the influence of hyperparameters and wavelet kernel functions on model results is also discussed.

Although the proposed model can achieve promising results in each case study, there are still some limitations that need to be addressed in future research. This includes the limitation that the extracted fault features can only be explained by expert knowledge if they have a clear fault mechanism (e.g., bearing and gear faults). However, the faults of other components are difficult to explain. Moreover, in such cases, it is also difficult to choose a proper kernel function. In future work, we plan to interpret the features in terms of impact components. Moreover, we intend to design a SGWN that has a learnable wavelet kernel function, thus avoiding the difficulty of choosing a suitable wavelet kernel function.

## APPENDIX
### DERIVATION PROCESS FOR APPROXIMATING SGWT OPERATOR

The interval for the standard Chebyshev polynomial is [-1, 1], however, for the scaling kernel function and the wavelet kernel function, their fitting interval is [0, $\lambda_{max}$]. Therefore, we can change the variables $y = \lambda_{max}(t+1)/2$, which transforms the interval [-1, 1] to [0, $\lambda_{max}$]. After that, we can obtain the formula of the shifted Chebyshev polynomial as

$$\bar{T}_k(y) = T_k(\frac{2y - \lambda_{max}}{\lambda_{max}}) = T_k(\frac{2y}{\lambda_{max}} - 1) \qquad (23)$$

With the above formula, we can get the first term and second term of the shifted Chebyshev polynomial are $\bar{T}_0(y) = 1$ and $\bar{T}_1(y) = 2y/\lambda_{max} - 1$, respectively. And the recurrence relation for calculating $\bar{T}_k(y)$ is $\bar{T}_k(y) = 2\left(\frac{2y}{\lambda_{max}} - 1\right)\bar{T}_{k-1}(y) - \bar{T}_{k-2}(y)$, so the recurrence relation of shifted Chebyshev polynomial for $h(\mathbf{L})$ and $g(a_j\mathbf{L})$ can be immediately implied as

$$\bar{T}_k(\mathbf{L}) = 2(\frac{2y}{\lambda_{max}}\mathbf{L} - \mathbf{I})\bar{T}_{k-1}(\mathbf{L}) - \bar{T}_{k-2}(\mathbf{L}) \qquad (24)$$

Therefore, the wavelet kernel function $g(a_j\mathbf{L})$ for each scale $a_j$ can be calculated by $g(a_j\mathbf{L}) = \frac{1}{2}\bar{c}_{j,0} + \sum_{k=1}^{K-1} \bar{c}_{j,k}\bar{T}_k(\mathbf{L})$ with $\bar{c}_{j,k} = \frac{2}{\pi}\int_0^\pi \cos(k\theta) g(\frac{a_j\lambda_{max}(\cos(\theta)+1)}{2}) d\theta$, while the scaling kernel function $h(\mathbf{L})$ is obtained by $h(\mathbf{L}) = \frac{1}{2}\bar{c}_{0,0} + \sum_{k=1}^{K-1} \bar{c}_{0,k}\bar{T}_k(\mathbf{L})$ with $\bar{c}_{0,k} = \frac{2}{\pi}\int_0^\pi \cos(k\theta) h(\frac{\lambda_{max}(\cos(\theta)+1)}{2}) d\theta$.

Finally, the SGWT operator $\mathbf{W} = [h(\mathbf{L}), g(a_1\mathbf{L}),...,g(a_J\mathbf{L})]^T$ can be obtained by calculating the approximated coefficient of $h(\mathbf{L})$ and a group of $g(a_j\mathbf{L})$ with $1 \le j \le J$.

## REFERENCES

[1] J. Cheng, Y. Yang, Z. Wu, H. Shao, H. Pan, and J. Cheng, "Ramanujan Fourier mode decomposition and its application in gear fault diagnosis," *IEEE Transactions on Industrial Informatics,* 2021.
[2] Q. Zhu *et al.*, "Real-time Defect Detection of Die Cast Rotor in Induction Motor Based on Circular Flux Sensing Coils," *IEEE Transactions on Industrial Informatics,* 2021.
[3] Y. Qin, C. Yuen, Y. Shao, B. Qin, and X. Li, "Slow-Varying Dynamics-Assisted Temporal Capsule Network for Machinery Remaining Useful Life Estimation," *IEEE Transactions on Cybernetics,* 2022.
[4] S. Zhang, Q. He, H. Zhang, and K. Ouyang, "Doppler correction using short-time MUSIC and angle interpolation resampling for wayside acoustic defective bearing diagnosis," *IEEE Transactions on Instrumentation and Measurement,* vol. 66, no. 4, pp. 671-680, 2017.
[5] T. Peng, C. Shen, S. Sun, and D. Wang, "Fault Feature Extractor based on Bootstrap Your Own Latent and Data Augmentation Algorithm for Unlabeled Vibration Signals," *IEEE Transactions on Industrial Electronics,* 2021.
[6] Y. Qin, Q. Qian, J. Luo, and H. Pu, "Deep Joint Distribution Alignment: A Novel Enhanced-Domain Adaptation Mechanism for Fault Transfer Diagnosis," *IEEE Transactions on Cybernetics,* 2022.
[7] H. Shao, H. Jiang, H. Zhang, and T. Liang, "Electric locomotive bearing fault diagnosis using a novel convolutional deep belief network," *IEEE Transactions on Industrial Electronics,* vol. 65, no. 3, pp. 2727-2736, 2017.
[8] Z. Chai, C. Zhao, and B. Huang, "Multisource-refined transfer network for industrial fault diagnosis under domain and category inconsistencies," *IEEE Transactions on Cybernetics,* vol. 52, no. 9, pp. 9784-9796, 2021.
[9] D. Zhu, X. Cheng, L. Yang, Y. Chen, and S. X. Yang, "Information fusion fault diagnosis method for deep-sea human occupied vehicle thruster

based on deep belief network," *IEEE Transactions on Cybernetics,* vol. 52, no. 9, pp. 9414-9427, 2021.

[10] J. Li, R. Huang, G. He, Y. Liao, Z. Wang, and W. Li, "A two-stage transfer adversarial network for intelligent fault diagnosis of rotating machinery with multiple new faults," *IEEE/ASME Transactions on Mechatronics,* vol. 26, no. 3, pp. 1591-1601, 2020.

[11] H. Wang, S. Li, L. Song, L. Cui, and P. Wang, "An enhanced intelligent diagnosis method based on multi-sensor image fusion via improved deep learning network," *IEEE Transactions on Instrumentation and measurement,* vol. 69, no. 6, pp. 2648-2657, 2019.

[12] D. Huang, W.-A. Zhang, F. Guo, W. Liu, and X. Shi, "Wavelet Packet Decomposition-Based Multiscale CNN for Fault Diagnosis of Wind Turbine Gearbox," *IEEE Transactions on Cybernetics,* 2021.

[13] X. Yu, B. Tang, and K. Zhang, "Fault Diagnosis of Wind Turbine Gearbox Using a Novel Method of Fast Deep Graph Convolutional Networks," *IEEE Transactions on Instrumentation and Measurement* vol. 70, p. 6502714, 2021.

[14] Z. Wu, S. Pan, F. Chen, G. Long, C. Zhang, and S. Y. Philip, "A comprehensive survey on graph neural networks," *IEEE transactions on neural networks and learning systems,* vol. 32, no. 1, pp. 4-24, 2020.

[15] Y. Li, D. Tarlow, M. Brockschmidt, and R. Zemel, "Gated graph sequence neural networks," *arXiv preprint arXiv:1511.05493,* 2015.

[16] W. Hamilton, Z. Ying, and J. Leskovec, "Inductive representation learning on large graphs," *Advances in neural information processing systems,* vol. 30, 2017.

[17] P. Veličković, G. Cucurull, A. Casanova, A. Romero, P. Lio, and Y. Bengio, "Graph attention networks," *arXiv preprint arXiv:1710.10903,* 2017.

[18] K. Xu, W. Hu, J. Leskovec, and S. Jegelka, "How powerful are graph neural networks?," *arXiv preprint arXiv:1810.00826,* 2018.

[19] M. Defferrard, X. Bresson, and P. Vandergheynst, "Convolutional neural networks on graphs with fast localized spectral filtering," *Advances in neural information processing systems,* vol. 29, 2016.

[20] T. N. Kipf and M. Welling, "Semi-supervised classification with graph convolutional networks," *arXiv preprint arXiv:1609.02907,* 2016.

[21] F. Wu, A. Souza, T. Zhang, C. Fifty, T. Yu, and K. Weinberger, "Simplifying graph convolutional networks," in *International conference on machine learning*, 2019: PMLR, pp. 6861-6871.

[22] T. Li, Z. Zhou, S. Li, C. Sun, R. Yan, and X. Chen, "The emerging graph neural networks for intelligent fault diagnostics and prognostics: A guideline and a benchmark study," *Mechanical Systems and Signal Processing,* vol. 168, p. 108653, 2022.

[23] Z. Chen, J. Xu, T. Peng, and C. Yang, "Graph convolutional network-based method for fault diagnosis using a hybrid of measurement and prior knowledge," *IEEE Transactions on Cybernetics,* 2021.

[24] X. Zhao, M. Jia, and Z. Liu, "Semisupervised graph convolution deep belief network for fault diagnosis of electormechanical system with limited labeled data," *IEEE Transactions on Industrial Informatics,* vol. 17, no. 8, pp. 5450-5460, 2020.

[25] K. Zhou, C. Yang, J. Liu, and Q. Xu, "Dynamic graph-based feature learning with few edges considering noisy samples for rotating machinery fault diagnosis," *IEEE Transactions on Industrial Electronics,* 2021.

[26] B. Zhao, X. Zhang, Z. Zhan, Q. Wu, and H. Zhang, "Multi-scale Graph-guided Convolutional Network with Node Attention for Intelligent Health State Diagnosis of a 3-PRR Planar Parallel Manipulator," *IEEE Transactions on Industrial Electronics,* 2021.

[27] K. Liu, N. Lu, F. Wu, R. Zhang, and F. Gao, "Model Fusion and Multiscale Feature Learning for Fault Diagnosis of Industrial Processes," *IEEE Transactions on Cybernetics,* 2022.

[28] S. Abu-El-Haija *et al.*, "Mixhop: Higher-order graph convolutional architectures via sparsified neighborhood mixing," in *international conference on machine learning*, 2019: PMLR, pp. 21-29.

[29] Q. Li, Z. Han, and X.-M. Wu, "Deeper insights into graph convolutional networks for semi-supervised learning," in *Thirty-Second AAAI conference on artificial intelligence*, 2018.

[30] T. Li, Z. Zhao, C. Sun, R. Yan, and X. Chen, "Multireceptive field graph convolutional networks for machine fault diagnosis," *IEEE Transactions on Industrial Electronics,* vol. 68, no. 12, pp. 12739-12749, 2021.

[31] S. Guo *et al.*, "Machine learning for metal additive manufacturing: Towards a physics-informed data-driven paradigm," *Journal of Manufacturing Systems,* vol. 62, pp. 145-163, 2022.

[32] T. Li *et al.*, "WaveletKernelNet: An interpretable deep neural network for industrial intelligent diagnosis," *IEEE Transactions on Systems, Man, and Cybernetics: Systems,* vol. 52, no. 4, pp. 2302-2312, 2021.

[33] T. Li, C. Sun, S. Li, Z. Wang, X. Chen, and R. Yan, "Explainable Graph Wavelet Denoising Network for Intelligent Fault Diagnosis," *IEEE Transactions on Neural Networks and Learning Systems,* 2022, doi: 10.1109/TNNLS.2022.3230458.

[34] D. Wang, Y. Chen, C. Shen, J. Zhong, Z. Peng, and C. Li, "Fully interpretable neural network for locating resonance frequency bands for machine condition monitoring," *Mechanical Systems and Signal Processing,* vol. 168, p. 108673, 2022.

[35] G. Xin *et al.*, "Fault Diagnosis of Wheelset Bearings in High-Speed Trains Using Logarithmic Short-time Fourier Transform and Modified Self-calibrated Residual Network," *IEEE Transactions on Industrial Informatics,* 2021.

[36] G. Michau, G. Frusque, and O. Fink, "Fully learnable deep wavelet transform for unsupervised monitoring of high-frequency time series," *Proceedings of the National Academy of Sciences,* vol. 119, no. 8, p. e2106598119, 2022.

[37] F. Gaëtan and F. Olga, "Learnable wavelet packet transform for data-adapted spectrograms," in *ICASSP 2022-2022 IEEE International Conference on Acoustics, Speech and Signal Processing (ICASSP)*, 2022: IEEE, pp. 3119-3123.

[38] D. K. Hammond, P. Vandergheynst, and R. Gribonval, "Wavelets on graphs via spectral graph theory," *Applied and Computational Harmonic Analysis,* vol. 30, no. 2, pp. 129-150, 2011.

[39] D. Thanou, D. I. Shuman, and P. Frossard, "Learning parametric dictionaries for signals on graphs," *IEEE Transactions on Signal Processing,* vol. 62, no. 15, pp. 3849-3862, 2014.

[40] X. Dong, G. Li, Y. Jia, B. Li, and K. He, "Non-iterative denoising algorithm for mechanical vibration signal using spectral graph wavelet transform and detrended fluctuation analysis," *Mechanical Systems and Signal Processing,* vol. 149, p. 107202, 2021.

[41] Z. Zhao *et al.*, "Deep learning algorithms for rotating machinery intelligent diagnosis: An open source benchmark study," *ISA transactions,* vol. 107, pp. 224-255, 2020.

[42] A. Vaswani *et al.*, "Attention is all you need," *Advances in neural information processing systems,* vol. 30, 2017.

[43] D. Wang, X. Zhao, L.-L. Kou, Y. Qin, Y. Zhao, and K.-L. Tsui, "A simple and fast guideline for generating enhanced/squared envelope spectra from spectral coherence for bearing fault diagnosis," *Mechanical Systems and Signal Processing,* vol. 122, pp. 754-768, 2019.

[44] J. Gilmer, S. S. Schoenholz, P. F. Riley, O. Vinyals, and G. E. Dahl, "Neural message passing for quantum chemistry," in *International conference on machine learning*, 2017: PMLR, pp. 1263-1272.

[45] C. Donnat, M. Zitnik, D. Hallac, and J. Leskovec, "Learning structural node embeddings via diffusion wavelets," in *Proceedings of the 24th ACM SIGKDD International Conference on Knowledge Discovery & Data Mining*, 2018, pp. 1320-1329.

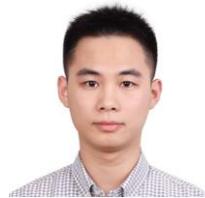

**Tianfu Li,** received the B.S. degree in Mechanical Engineering from Chongqing University, China in 2018. He is currently working towards the Ph.D. degree in mechanical engineering in the Department of Mechanical Engineering, Xi'an Jiaotong University, Xi'an, China. From 2022 to 2023, he is a visiting Ph.D. student at the laboratory of Intelligent Maintenance and Operations Systems, EPFL, Lausanne, Switzerland.

His current research is focused on graph representation learning, explainable artificial intelligence, and intelligent maintenance.

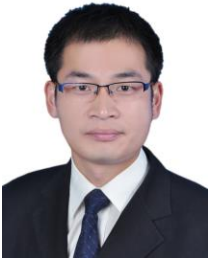

**Chuang Sun**, received the Ph.D. degree in mechanical engineering from Xi'an Jiaotong University, Xi'an, China, in 2014. From Mar. 2015 to Mar. 2016, he was a postdoc at Case Western Reserve University, USA. He is now a Professor in School of Mechanical Engineering at Xi'an Jiaotong University.

His research is focused on manifold learning, deep learning, sparse representation, mechanical fault diagnosis and prognosis, remaining useful life prediction.

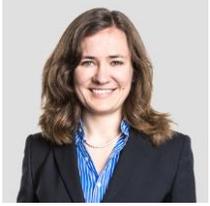

**Olga Fink**, received her Ph.D. degree from ETH Zurich in 2014 and a Diploma degree in industrial engineering from the Hamburg University of Technology, Hamburg, Germany, in 2008. She has been assistant professor of intelligent maintenance and operations systems at EPFL since March 2022. Before joining EPFL faculty, Olga was assistant professor of intelligent maintenance systems at ETH Zurich from 2018 to 2022, being awarded the prestigious professorship grant of the Swiss National Science Foundation (SNSF). Between 2014 and 2018, she was heading the research group "Smart Maintenance" at the Zurich University of Applied Sciences (ZHAW). And Olga is also a research affiliate at Massachusetts Institute of Technology.

Her research focuses on Hybrid Algorithms Fusing Physics-Based Models and Deep Learning Algorithms, Hybrid Operational Digital Twins, Transfer Learning, Self-Supervised Learning, Deep Reinforcement Learning and Multi-Agent Systems for Intelligent Maintenance and Operations of Infrastructure and Complex Assets.

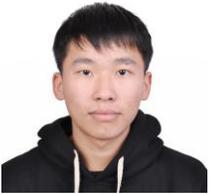

**Yuangui Yang,** received the B.S. degree in Mechanical Engineering from Xi'an Jiaotong University, China in 2020. He is currently working towards the Ph.D. degree in mechanical engineering in the Department of Mechanical Engineering, Xi'an Jiaotong University, Xi'an, China.

His current research is focused on explainable deep learning, graph neural network, and gearbox fault diagnosis.

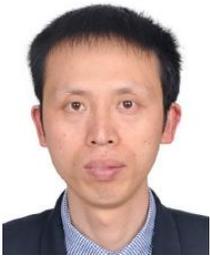

**Xuefeng Chen,** (M'12) received the Ph.D. degree from Xi'an Jiaotong University, Xi'an, China, in 2004. He is currently a Professor of Mechanical Engineering with Xi'an Jiaotong University.

His current research interests include finite-element method, mechanical system and signal processing, diagnosis and prognosis for complicated industrial systems, smart structures, aero-engine fault diagnosis, and wind turbine system monitoring.

Dr. Chen was a recipient of the National Excellent Doctoral Dissertation of China in 2007, the Second Award of Technology Invention of China in 2009, the National Science Fund for Distinguished Young Scholars in 2012, and a Chief Scientist of the National Key Basic Research Program of China (973 Program) in 2015. He is the Chapter Chairman of the IEEE Xi'an and Chengdu Joint Section Instrumentation and Measurement Society.

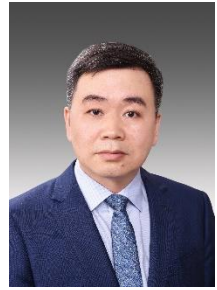

**Ruqiang Yan** (M'07, SM'11) received the M.S. degree in precision instrument and machinery from the University of Science and Technology of China, Hefei, China, in 2002, and the Ph.D. degree in mechanical engineering from the University of Massachusetts at Amherst, Amherst, MA, USA, in 2007.

From 2009 to 2018, he was a Professor with the School of Instrument Science and Engineering, Southeast University, Nanjing, China. He joined the School of Mechanical Engineering, Xi'an Jiaotong University, Xi'an, China, in 2018. His research interests include data analytics, machine learning, and energy-efficient sensing and sensor networks for the condition monitoring and health diagnosis of large-scale, complex, dynamical systems.

Dr. Yan is a Fellow of ASME (2019). His honors and awards include the IEEE Instrumentation and Measurement Society Technical Award in 2019, the New Century Excellent Talents in University Award from the Ministry of Education in China in 2009, and multiple best paper awards. He is also the Associate Editor-in-Chief of the IEEE Transactions on Instrumentation and Measurement and an Associate Editor of the IEEE Systems Journal and the IEEE Sensors Journal.